\documentclass[12pt,manuscript]{elsart}
\usepackage{graphicx,harvard,amssymb}
\input{psfig}

\def\spjetp{Sov.Phys.JETP}
\def\casp{Comments.Astrophys.Space Phys.}

\def\physrep{Phys.Rep.}
\def\araa{Annu. Rev. Astron. Astrophys.}

\def\apj{ApJ.}

\def\mnras{MNRAS}

\def\ea{\ et al. \,}

\def\be{\begin{equation}}
\def\ee{\end{equation}}
\def\rel{relativistic\,}
\def\nrel{nonrelativistic\,}
\def\sz{Sunyaev \& Zeldovich\,}
                                
\def\gs{\mathrel{\raise0.35ex\hbox{$\scriptstyle >$}\kern-0.6em 
\lower0.40ex\hbox{{$\scriptstyle \sim$}}}}
\def\ls{\mathrel{\raise0.35ex\hbox{$\scriptstyle <$}\kern-0.6em 
\lower0.40ex\hbox{{$\scriptstyle \sim$}}}}

\makeatletter
\def\elsartstyle{%
        \def\normalsize{\@setfontsize\normalsize\@xiipt{14.5}}
        \def\small{\@setfontsize\small\@xipt{13.6}}
        \let\footnotesize=\small
        \def\large{\@setfontsize\large\@xivpt{18}}
        \def\Large{\@setfontsize\Large\@xviipt{22}}
        \skip\@mpfootins = 18\p@ \@plus 2\p@
        \normalsize
}
\makeatother

\def\url#1{{\ttfamily\def\/{/\discretionary{}{}{}}#1}}

\begin{document}

\begin{frontmatter}
\title{Quantitative Description of the Sunyaev-Zeldovich\\ 
Effect: Analytic Approximations} 

\author{Meir Shimon\thanksref{email}}
\address{School of Physics and Astronomy, Tel Aviv University, Tel 
Aviv, 69978, Israel}

\thanks[email]{E-mail: meirs@ccsg.tau.ac.il}

\and

\author{Yoel Rephaeli}

\address{School of Physics and Astronomy, Tel Aviv University, Tel
Aviv, 69978, Israel, \\and\\ Center for Astrophysics and Space
Sciences, University of California, San Diego, La Jolla,
CA\,92093}

\begin{abstract}

Various aspects of relativistic calculations of the 
Sunyaev-Zeldovich effect are explored and clarified. 
We first formally show that the main previous approaches 
to the calculation of the relativistically generalized 
thermal component of the effect are equivalent. Our 
detailed description of the full effect results in a 
somewhat improved formulation. Analytic approximations 
to the exact calculation of the change of the photon 
occupation number in the scattering, $\Delta n$, 
are extended to powers of the gas temperature and cluster 
velocity that are higher than in similar published 
treatments. For the purely thermal and purely kinematic 
components, we obtain identical terms up to the highest 
common orders in temperature and cluster velocity, and 
to second order in the Thomson optical depth, as reported 
in previous treatments, but we get slightly different 
expressions for the terms that depend on both the gas 
temperature and cluster velocity. We also obtain an 
accurate expression for the crossover frequency.

\end{abstract}

\end{frontmatter}

\section{Introduction}

Compton scattering of the cosmic microwave background (CMB) by 
electrons in the hot gas in clusters of galaxies (\sz 1972) 
imprints on the radiation a spectral signature -- the 
Sunyaev-Zeldovich (S-Z) effect -- that has long been recognized as 
a uniquely important feature which can be used as an indispensable 
cosmological probe. Measurements with single-dish radio telescopes 
and interferometric arrays have led so far to the detection and 
imaging of the effect in more than 60 clusters. These measurements 
yield directly the properties of the hot intracluster (IC) gas, 
and indirectly the {\it total} mass of the cluster, insight on the 
evolution of clusters, and the values of the basic cosmological 
parameters (for reviews, see Rephaeli 1995a, Birkinshaw 1999, 
Carlstrom \ea 2002). 

Exact description of the effect is needed in order to extract 
precise information from current measurements, and especially 
from upcoming high-frequency observations. The calculations of 
Sunyaev \& Zeldovich (1972) were based on a solution to the 
Kompaneets (1957) equation, a {\it \nrel} diffusion approximation 
to the exact kinetic (Boltzmann) equation that describes the 
scattering. The result of their treatment is a simple expression 
for the intensity change resulting from scattering of the CMB by 
electrons with a thermal velocity distribution. The effect has a 
second kinematic (Doppler) component which is proportional to the 
component of the cluster peculiar velocity along the line of 
sight, $v_r$ (Sunyaev \& Zeldovich 1980). 

Rephaeli (1995b) has shown that the approximate nonrelativistic treatment
of Sunyaev \& Zeldovich (1972) is not sufficiently accurate for use 
of the effect as a precise cosmological probe. High electron 
velocities in the IC gas, and relatively large photon energy change 
in the scattering, necessitate a more exact \rel calculation. 
Using the exact probability distribution, and the relativistically
correct form of the electron Maxwellian velocity distribution,
Rephaeli calculated the resulting intensity change in the limit of 
small optical depth to Thomson scattering, $\tau$, keeping terms 
linear in $\tau$. This semi-analytic calculation demonstrated that 
the relativistic spectral distribution of the intensity change is 
appreciably different from that derived by Sunyaev \& Zeldovich 
(1972). Deviations from the \sz expression are especially large 
near the crossover frequency, where the purely thermal effect 
vanishes, and on the Wien side where the effect of boosting 
photons to relatively high frequencies by energetic electrons 
is important.
         
The results of the calculations of Rephaeli (1995b) generated 
considerable interest that led to various generalizations and 
extensions of the relativistic treatment. Challinor \& Lasenby 
(1998) obtained an analytic approximation to the solution of the 
relativistically generalized Kompaneets equation. Nozawa \ea (1998a) 
improved the accuracy of this approximation by expanding to fifth 
order in $\Theta= kT_{e}/m_{e}c^{2}$, where $T_e$ is the electron 
temperature. Sazonov \& Sunyaev (1998) and Nozawa et al. (1998b) 
have extended the relativistic treatment also to the kinematic 
component obtaining -- for the first time -- the leading cross 
terms in the expression for the combined (thermal and kinematic) 
intensity change, $\Delta I_t + \Delta I_k$, which depends on both 
$T_e$ and $v_{c}$, the cluster peculiar velocity. An analytic fit 
to the numerical solution, valid for $0.02 \leq \Theta_{e} 
\leq 0.05$, and $x \equiv h\nu/kT \leq 20$ (for $\nu \leq 1130$ 
GHz; $T$ is the CMB temperature), has been given by Nozawa et 
al. (2000). 

Cluster X-ray and S-Z measurements have significantly improved in 
the last few years. Many nearby and distant clusters have been 
sensitively mapped spectrally and spatially by the {\it Chandra} 
and XMM satellites, yielding a wealth of information on the 
complex morphology and thermal structure of IC gas. Interferometric 
S-Z measurements, and upcoming high resolution observations with 
bolometric multi-frequency arrays containing large number of 
elements, expand the scope of S-Z science and further motivate 
more accurate treatment of the basic effect and its realistic 
modeling. 

In this paper we clarify various aspects of a relativistically exact
description of the S-Z effect, and derive accurate analytic
expressions for the intensity change and crossover frequency that 
are valid over wide regions of parameter space. Several different 
approaches have been adopted in the relativistic calculation of 
the effect, and although the approximate analytic expressions that 
were derived in these treatments are either consistent when 
compared at the same level of accuracy, or yield consistent 
numerical results, it has not yet been formally shown that the 
three main approaches to \rel calculation of the S-Z effect are 
indeed equivalent. This equivalence is formally proved in Section 
2.1. In order to provide more accurate analytic approximation to 
the thermal component of the effect, we derive -- in section 
2.2 -- an analytic fit to the thermal component of the effect 
which is accurate to orders $\tau^{1}\Theta^{8}$. In view 
of the possibility that in some rich clusters $\tau$ is as high as 
$0.03$, the approximate analytic expansion to first order in $\tau$ 
and to eighth order in $\Theta$ necessitates also the inclusion of 
multiple scatterings, of order $\tau^2$. This has been considered 
by Itoh et al. (2001), and is treated further here in section 2.2, 
where we consider the contribution of two successive scatterings 
(terms to order $\tau^{2}$) and find that the $\tau^{2}\Theta^{2}$ 
-- $\tau^{2}\Theta^{5}$ contributions are appreciable near the 
crossover frequency for which we derive an analytic approximation. 
In section 2.2 we also include an analytic expression for the 
change in the photon occupation number that is accurate for gas 
temperatures as high as 50 keV. In sections 2.3 and 2.4 the 
kinematic and the full effect are calculated, including the cross 
terms that depend on both the gas temperature and the cluster 
velocity. Our results for these terms differ somewhat from those 
obtained by other groups. The last section (3) contains a more 
general discussion of the results presented in this paper.

\section{Exact Description of the S-Z Effect}

In this section we explicitly calculate the thermal, kinematic and
cross terms of the S-Z effect. While such calculations have 
already been carried out in various papers, our approach here is 
more direct, and the results are roughly at the same level of 
accuracy as achieved in the recent work of Itoh \& Nozawa (2003), 
where the results of a numerical fit are given in a tabulated 
form. Following a detailed description of the full effect, and our 
analytic approximations, we present a fit to the exact calculation 
that is valid at gas temperature well beyond 15 keV. We begin with 
a general comparison of the main theoretical descriptions of the 
effect.

\subsection{Equivalent Relativistic Treatments}

Three main approaches have been adopted in the calculation of the 
relativistic S-Z effect, those employed by Rephaeli (1995a), 
Sazonov \& Sunyaev (1998), and Nozawa et al. (1998a). Different 
analytic approximations to the exact expressions that were obtained 
in these treatments were found to be consistent when compared to 
the highest common order of $\Theta$. Here we formally prove that 
these three approaches are indeed equivalent. Basically, this 
follows from the fact that although the method used by Nozawa et 
al. (1998a) is more general than those used by Rephaeli (1995a, 
1995b) and Sazonov \& Sunyaev (1998), the simplifying assumptions 
made in deriving analytic approximations, namely neglecting 
electron recoil, and expansion in powers of a small quantity -- 
the ratio of the change in the photon energy to the photon energy 
-- render it equivalent to the other two approaches.

Consider first the method employed by Rephaeli (1995a, 1995b).
The probability of scattering of an incoming photon (direction
$\mu_{0}=\cos\theta_{0}$) to the direction 
$\mu'_{0}=\cos\theta'_{0}$ is (Chandrasekhar 1950)
\begin{eqnarray}
f\left(\mu_{0},\mu'_{0}\right)=\frac{3}{8}\left[1+
\mu_{0}^{2}\mu'^{2}_{0}+\frac{1}{2}\left(1-\mu_{0}^{2}\right)
\left(1-\mu'^{2}_{0}\right)\right]
\label{fmumu'}
\end{eqnarray}
where the subscript 0 refers to the electron rest frame. The 
frequency shift is
\begin{eqnarray}
s=\ln\left(\nu'/\nu\right)=\ln\left(\frac{1+\beta\mu'_{0}}{1+
\beta\mu_{0}}\right)
\label{s}
\end{eqnarray}
where $\nu$, $\nu'$ are the frequency of the photon before and 
after the scattering, and $\beta=v/c$ is the dimensionless electron 
velocity in the CMB frame. It is somewhat more convenient to use 
the variables $\mu,s$ and $\beta$ instead of $\mu,\mu'$ and 
$\beta$. The probability that a scattering results in a frequency 
shift $s$ is (Wright 1979)
\begin{eqnarray}
\mathcal{P}\left(s,\beta\right)=
\frac{1}{2\gamma^{4}\beta}\int\frac{e^{s}f\left(\mu_{0},
\mu'_{0}\right)}{\left(1+\beta\mu_{0}\right)^{2}}d\mu_{0}.
\label{P}
\end{eqnarray}
Averaging over a Maxwellian distribution for the electrons yields
\begin{eqnarray}
\mathcal{P}_{1}\left(s\right)=\frac{\int\beta^{2}\gamma^{5}e^{-
\frac{\left(\gamma-1\right)}{\Theta}}\mathcal{P}\left(s,
\beta\right)d\beta}{\int\beta^{2}\gamma^{5}e^{-\frac{\left(\gamma-
1\right)}{\Theta}}d\beta}.
\label{P1}
\end{eqnarray}
The total change in the photon occupation number along a line of 
sight (los) to the cluster, due to scattering off electrons with 
thermal ($t$) velocities, can now be written as
\begin{eqnarray}
\Delta
n_{t}(x)=\tau\int_{-\infty}^{\infty}\left[n\left(xe^{s}\right)-
n\left(x\right)\right]\mathcal{P}_{1}\left(s\right)ds
\label{deltant}
\end{eqnarray}
where $x$ is the dimensionless frequency, $x=h\nu/kT$, and 
$\tau\equiv\sigma_{T}\int n_{e}dl$ is the optical depth to Thomson 
scattering. 

In the approach of Sazonov \& Sunyaev (1998), which is similar to 
that of Rephaeli (1995a), the starting point is the photon transfer 
equation
\begin{eqnarray}
\frac{dn_{0}}{dt_{0}}(\mu_{0},\nu_{0})=
cn_{e0}\int\frac{d\sigma}{d\Omega'_{0}}\left[n_{0}\left(\mu'_{0},
\nu_{0}\right)-n_{0}\left(\mu_{0},\nu_{0}\right)\right]d\Omega'_{0},
\label{transferss}
\end{eqnarray}
where the Planckian occupation number is
\begin{eqnarray} 
n_{0}=\frac{1}{e^{x_{0}\gamma\left(1+\beta\mu_{0}\right)}-1}, 
\label{n0}
\end{eqnarray} 
and $n_{e0}$ \& $\mu_{0}$ are the electron number density and the 
cosine of the angle between the directions of the electron motion, 
and the photon velocity as seen in this frame, respectively. The 
differential scattering cross section in the electron rest frame is
\begin{eqnarray}
\frac{d\sigma}{d\Omega'_{0}}=\frac{1}{2}\sigma_{T}
f(\mu_{0},\mu'_{0}), 
\label{diffcross}
\end{eqnarray}
where $\sigma_{T}$ is the Thomson cross section. Lorentz 
transformation of equation (\ref{transferss}) with $n_{e0}
=n_{e}/\gamma$ leads to 
\begin{eqnarray}
\frac{dn_{0}}{dt}(\mu_{0},\nu_{0})=\int\frac{cn_{e}}{\gamma^{2}
\left(1+\beta\mu_{0}\right)}\delta n\frac{d\sigma}
{d\Omega'_{0}} d\Omega'_{0},
\label{9}
\end{eqnarray}
where
\begin{eqnarray}
\delta n\equiv
n_{0}\left(\mu'_{0},\nu_{0}\right)-n_{0}\left(\mu_{0},\nu_{0}\right)
=\frac{1}{e^{xz}-1}-\frac{1}{e^{x}-1}.
\label{10}
\end{eqnarray}
Now, in the Thomson limit, 
\begin{eqnarray}
z&=&\frac{1+\beta\mu'_{0}}{1+\beta\mu_{0}},
\label{z}
\end{eqnarray}
and
\begin{eqnarray}
\frac{dn_{t}}{d\tau}=\int\int\int\frac{d\sigma}
{d\Omega'_{0}}\frac{\delta n}{\left(1+\beta\mu_{0}\right)^{3}}
e^{-\frac{\gamma-1}{\Theta}}d\mu_{0}d\mu'_{0}\gamma\beta^{2}d\beta/D
\label{12}
\end{eqnarray}
where
\begin{eqnarray}
D=\int_{0}^{1} e^{-\frac{\gamma -1}
{\Theta}}\gamma^{5}\beta^{2}d\beta.
\label{D}
\end{eqnarray}
This is exactly the expression that was derived by Rephaeli (1995a, 
1995b) using a different approach, hence the two methods are shown 
to be fully equivalent.

In the third method, that of Nozawa et al. (1998a), the starting 
point is a more basic expression for the time rate of change of the 
photon occupation number given by a generalized Kompaneets equation 
\begin{eqnarray}
\frac{\partial n}{\partial
  t}\left(\omega\right) &=&
-2\int\int\int\frac{d^{3}p}{\left(2\pi\right)^{3}}d^{3}p'd^{3}k'
W\left[n\left(\omega\right)\left(1+n\left(\omega'\right)\right)
P_{e}\left(E\right)\right.\nonumber\\
&-&\left.n\left(\omega'\right)\left(1+n\left(\omega\right)\right)
P_{e}\left(E'\right)\right],
\label{transitoh}
\end{eqnarray}
where
\begin{eqnarray}
W &\equiv &
\left(\frac{e^{2}}{4\pi}\right)^{2}\frac{X}{2\omega\omega'EE'}
\delta^{4}\left(p+k-p'-k'\right),\label{15}\\
X &\equiv &
-\left(\frac{\rho}{\rho'}+\frac{\rho'}{\rho}\right)+
4m^{4}\left(\frac{\rho}{\rho'}+\frac{\rho'}{\rho}\right)^{2}-
4m^{2}\left(\frac{1}{\rho}+\frac{1}{\rho'}\right),\label{16}\\
\rho &\equiv & -2\left(p\cdot k\right)=-2\omega
E\left(1-\beta\mu_{0}\right),\label{17}\\
\rho' &\equiv & 2\left(p\cdot k'\right)=2\omega'
E\left(1-\beta\mu'_{0}\right)\label{18},
\end{eqnarray}
$p$ and $k$ ($p'$ and $k'$) are the ingoing (outgoing) electron and
photon 4-momentum
($p=(E,\vec{p})$, $k=(\omega,\vec{k})$, $p'=(E',\vec{p'})$, $k'=(\omega',\vec{k'})$),
and
$\mu_{0}$ ($\mu'_{0}$) is the cosine angle between $\vec{k}$ and
$\vec{p}$ ($\vec{k'}$ and $\vec{p'}$) in the electron rest frame.
$m$ and $e$ are the electron mass and electric charge and $P_{e}(E)$ is
the electron energy distribution.

In the Thomson limit, $\omega'_{0}=\omega_{0}$, therefore $E=E'$, 
and Equation (14) simplifies to
\begin{eqnarray}
\frac{\partial n}{\partial t}\left(x\right) &=&
-2\int\frac{d^{3}p}{\left(2\pi\right)^{3}}d^{3}k'W
\left[n\left(x\right)-n\left(x'\right)\right]P_{e}\left(E\right).
\label{19}
\end{eqnarray}
Now $X$ is essentially the differential cross section, equation 
(8), used by Rephaeli (1995a) and Sazonov \& Sunyaev (1998) in a 
less general form. The approximations made by Nozawa et al. (1998a) 
are based on the fact that $\Delta x\ll 1$ in the CMB frame (which 
is in fact, used as a perturbation parameter), and that 
$T/T_{e}\ll 1$. However, since $h\nu/(m_{e}c^{2}\Theta)
=h\nu/kT_{e}\sim T/T_{e}\ll 1$, it follows that 
$h\nu/m_{e}c^{2}\ll\Theta$, and the Thomson limit is valid. 
Note that $\Delta x\ll 1$ implies that $\beta\ll 1$, and therefore 
the condition $h\nu/(m_{e}c^{2})\ll 1$ still holds in the electron
rest frame. It can now be seen that even though the treatment of 
Nozawa et al. (1998a) is seemingly more general, their final 
expression -- in the limit when the commonly adopted approximations 
are valid -- reduces to those obtained by Rephaeli (1995a) and 
Sazonov \& Sunyaev (1998). Therefore, all the three methods are 
equivalent and will therefore yield equivalent results to all 
orders in $\Theta$, provided the expansion in $\Theta$ is valid 
($\Theta\ll 1$).

\subsection{The Thermal Component}

\subsubsection{Single Scattering Limit}

In an optically very thin medium with $\tau \leq 0.01$, only 
single scatterings of CMB photons need to be (practically) 
considered, and the leading term in the expression for $\Delta n$ 
is (by far) of order $\tau$. Using equation (\ref{9}) with an 
integral over (relativistically generalized expression for) a 
Maxwellian distribution of electron velocities, we obtain
\begin{eqnarray}
\frac{dn_{t}}{dt}&=&\frac{1}{D}\int_{\beta=0}^{1}\int_{\mu'_{0}
=-1}^{1}\int_{\mu_{0}=-1}^{1}\frac{cn_{e}}
{\gamma^{4}\left(1+\beta\mu_{0}\right)^{3}}\frac{d\sigma}
{d\Omega'_{0}}\delta n e^{-\frac{\gamma-1}{\Theta}}\gamma^{5}
\beta^{2}d\beta d\mu'_{0}d\mu_{0}.
\label{20}
\end{eqnarray}
To first order in $\tau$, $\Delta n$ is
\begin{eqnarray}
\Delta
n_{t}&=&\frac{\tau}{2D}\int_{\beta=0}^{1}\int_{\mu'_{0}
=-1}^{1}\int_{\mu_{0}=-1}^{1}\frac{\delta
  n}{\gamma^{4}\left(1+\beta\mu_{0}\right)^{3}}
f\left(\mu_{0},\mu'_{0}\right)e^{-\frac{\gamma-
1}{\Theta}}\gamma^{5}\beta^{2}d\beta d\mu'_{0}d\mu_{0}.
\label{21}
\end{eqnarray}
To perform the integrations in this expression, we expand the
integrand in powers of $\beta$ around $\beta =0$ with the 
derivatives (at $\beta=0$) compactly written by defining
\begin{eqnarray}
A_{n}=-2^{n-2}x^{n}[\sinh^{-2}\frac{x}{2}]^{(n-1)},
\label{An}
\end{eqnarray}
where $f^{(n)}$ denotes the n'th derivative of $f$. Carrying out the
expansion to $\beta^{16}$ and integrating yield 
\begin{eqnarray}
\frac{\Delta n_{t}}{\tau}&=&\Theta F_{1}+\Theta^{2}F_{2}+
\Theta^{3}F_{3}+\Theta^{4}F_{4}+\Theta^{5}F_{5}+\Theta^{6}F_{6}+\Theta^{7}F_{7}+\Theta^{8}F_{8}
\label{deltantau}
\end{eqnarray}
where $F_{i}$ are listed in Appendix A.

It is instructive to compare our result in equation 
(\ref{deltantau}) to equations (2.25)-(2.30) of Nozawa et al. 
(1998a) that were obtained using a different approach. The 
functions $A_{i}$ can be expressed in terms of $\tilde{X}$ and 
$\tilde{S}$
\begin{eqnarray}
\tilde{X}&\equiv & x\coth\frac{x}{2},\nonumber\\
\tilde{S}&\equiv &\frac{x}{\sinh\frac{x}{2}},
\label{XS}
\end{eqnarray}
and the relations between $A_{i}$, $\tilde{X}$ and $\tilde{S}$ are given in 
Appendix B. Substituting the expressions in Appendix B back into 
$F_{1}$ - $F_{5}$ and equation (\ref{XS}), we obtain exactly 
equations (2.25)-(2.30) of Nozawa et al. (1998a) term by term. 
This is, of course, just what we expect given the proven 
equivalence of the approaches discussed above. 

\subsubsection{Multiple Scatterings} 

Although formally very accurate, the single scattering analytic
approximation to order $\Theta^{8}$ is actually limited to small 
values of $\tau$. This optical depth is typically $\sim 0.01$ in 
a rich cluster, but values as high as of up to $0.03$ are not
unexpected in some clusters. Thus, the possibility of double 
scatterings is not ignorable to the level of accuracy afforded by 
the above expansion to order $\tau \Theta^{8}$. A more self 
consistent calculation must include the higher order terms
up to $\tau^{2}\Theta^{5}$, which can be comparable to the
$\tau \Theta^{8}$ term. The importance of considering the case 
of finite optical depth was already pointed out by Molnar \& 
Birkinshaw (1999), Itoh et al. (2001), and Dolgov et al. (2001). 
Our goal here is to extend the analytic approximation by 
calculating all the relevant contributions up to and including 
terms that are of order $\tau^{2}\Theta^{5}$. 

The impact of second scatterings on the occupation number can 
be calculated iteratively by describing the incident radiation in 
terms of the occupation number, $n_{1}$, that results after the 
radiation has already been singly scattered. In this way one can 
extend the treatment to the case of any number of scatterings 
(Wright 1979). Now, $n_{1}$ is simply 
\begin{eqnarray}
n_{1}=n_{0}+\Delta n_{0},
\label{25}
\end{eqnarray}
where $\Delta n_{0}$ is given in equation (\ref{deltantau}). 
Substituting $n_{1}$ into the right hand side of equation 
(\ref{12}) we can solve for $\Delta n_{1}$. Note that the result 
of the naive iteration should be divided by $2!$ to avoid double 
counting; a given total frequency shift ($\nu'/\nu=z$) results from 
two scatterings irrespective of the order of the two scattering 
events. The expression for the occupation number of the doubly 
scattered radiation is 
\begin{eqnarray}
n'=e^{-\tau}\left(n_{0}+\tau n_{0}\otimes \mathcal{P}+\frac{1}{2}
\tau^{2}n_{0}\otimes \mathcal{P}\otimes \mathcal{P}\right),
\label{26}
\end{eqnarray}
where $\otimes$ denotes convolution and (as before) $\mathcal{P}$ 
is the scattering probability, equation (\ref{P}). Expanding the 
exponent to second order in $\tau$, we have 
\begin{eqnarray}
\Delta n_{t}&=&\left(\tau-\frac{1}{2}\tau^{2}\right)\left(n\otimes
\mathcal{P} -n\right)+\frac{1}{2}\tau^{2}\left(n\otimes\mathcal{P}
\otimes\mathcal{P}-n\otimes\mathcal{P}\right).
\label{27}
\end{eqnarray}
The first term on the right hand side of equation (27)
($n\otimes\mathcal{P}-n$) was evaluated in section 2.2.1; here we 
calculate the $\tau^{2}$ contribution. Performing the 
integrations in a way similar to the calculation of the first 
order $\tau$ contribution, we obtain
\begin{eqnarray}
&&\frac{1}{2}\tau^{2}(n\otimes\mathcal{P}
\otimes\mathcal{P}-n\otimes
\mathcal{P})=\tau^{2}\Theta^{2}F_{9}+\tau^{2}\Theta^{3}F_{10}+\tau^{2}\Theta^{4}F_{11}\nonumber\\
&+&\tau^{2}\Theta^{5}F_{12}.
\label{28}
\end{eqnarray}
Altogether, the resulting change in the occupation number 
calculated to orders $\tau\Theta^{8}$ and $\tau^{2}\Theta^{5}$ is
\begin{eqnarray}
\Delta n_{t}&=&\tau(\Theta
F_{1}+\Theta^{2}F_{2}+\Theta^{3}F_{3}+\Theta^{4}F_{4}+
\Theta^{5}F_{5}+\Theta^{6}F_{6}+\Theta^{7}F_{7}\nonumber\\
&+&\Theta^{8}F_{8})+\tau^{2}(\Theta^{2}F_{9}+\Theta^{3}F_{10}+\Theta^{4}F_{11}+\Theta^{5}F_{12}).
\label{order9}
\end{eqnarray}
It can be easily shown that equation (29) satisfies photon number 
conservation. 

The $\tau^{2}$ terms are important especially near the crossover 
frequency, whose value depends both on $\Theta$ and $\tau$, as 
derived explicitly below. Also of interest is the unignorable 
effect of double scatterings at frequencies on the far Wien side 
of the spectrum, due to the fact that the total change in the 
frequency of the photon after two scatterings is particularly 
high. More specifically, we examined the contribution of these 
terms to $\Delta I$. For $\tau$ in the range [0.01,0.04] and 
$\Theta$ in the range [1/130, 1/35], the corrections due to 
these terms are larger than $1\%$ near the crossover frequency 
as expected. Clearly, the spectral band around this frequency 
-- for which the corrections are not ignorable -- widens as 
$\Theta$ increases. 

\subsubsection{Analytic Fit to the Thermal Effect}

The observed range of IC gas temperatures is $\sim 3-15$ keV, 
when the gas is taken to be isothermal (in the fit to the 
observed X-ray spectrum). High spatial and spectral resolution 
measurements with the {\it Chandra} and XMM satellites show more 
and more that the temperature distribution is non-uniform, 
which is only to be expected given the hierarchical growth of 
structure in the context of CDM models. A more realistic modeling 
of IC gas yields hotter and colder regions in some clusters, 
perhaps reflecting the nature of the formation process of clusters 
from aggregates or sub-clumps. In a naive modeling of the 
temperature evolution of clusters a virial mass-temperature 
relation is sometimes obtained with clusters of a given mass 
having a higher temperature the earlier they virialized. If so, 
a roughly linear relation would be predicted between the 
temperature and the virialization redshift of the cluster. 
Irrespective of the reality of such a scaling, it is of interest 
to extend the accuracy of the analytic approximations for 
$\Delta n$ to values of $kT_e$ which are higher than $15$ keV. 
Equation (\ref{deltantau}) is not sufficiently accurate at higher 
temperatures because the series expansion does not converge and 
rapidly oscillates around the exact numerical result (equation 
\ref{21}). 

Here we obtain a more accurate fit to the results 
from the exact calculation using a polynomial function of the form
\begin{eqnarray}
\Delta
n(x)/\tau=\sum_{i=1}^{5}\sum_{j=1}^{10}b_{ij}A_{j}\Theta^{i}
=\sum_{i=1}^{5}H_{i}\Theta^{i}
\label{fit}
\end{eqnarray}
where the numerical coefficients $b_{ij}$ are obtained by means 
of a best fit to the exact expression (equation \ref{21}). The 
coefficients $H_{i}$ are given in Appendix A. This expression is 
in agreement with the exact calculation to better than 2.5\% for 
$\Theta<1/10$, or $kT_e < 50$ keV. This level of 
accuracy is higher than that obtained by Itoh et al. (2001) who 
used the analytic expansion (equation \ref{deltantau}). Evidently, 
higher accuracy is attained by not insisting on using the exact 
form of the analytic expansion also at higher temperatures; our 
approach is thus functionally less constrained. 
Very recently, Itoh \& Nozawa (2003) have extended their 
previous analytic approximation for $\Delta n$ to higher 
temperatures by fitting the results of the exact numerical 
calculation with a polynomial of order $\Theta^5$ for $x<1.2$, 
and a fully numerical fit to a polynomial in powers of both 
$\Theta$ (to 13th order) and $x$ for $x>1.2$ (with the large 
array of coefficients given in a tabulated form). This more 
cumbersome procedure does result in a more accurate representation 
of the exact numerical calculation, which is generally within 
$0.4\%$.

\subsubsection{Crossover Frequency}

As was pointed out in Rephaeli (1995b), Sazonov \& Sunyaev (1998)
and Nozawa et al. (1998a), the crossover frequency ($x_{0}$)
depends on the gas temperature. In previously derived expressions
for the functional form of $x_{0}$, only the dependence on
$\Theta$ was considered, and deduced from a $\chi^{2}$ fit.
In order to find an analytic approximation to the shifted
crossover frequency we write
\begin{eqnarray}
x_{0}&=&3.8300161\left(1+c_{1}\Theta+c_{2}\Theta^{2}+
c_{3}\Theta^{3} +c_{4}\Theta^{4}+c_{5}\tau+c_{6}\tau\Theta\right.\nonumber\\
&+&\left.c_{7}\tau\Theta^{2}\right),
\end{eqnarray}
and substitute this expression in equation (29). Expanding in the
small parameters $\tau$ and $\Theta$, we then obtain the values
of the coefficients in the expression for $x_{0}$ by setting
$\Delta n_{t}=0$,
\begin{eqnarray}
x_{0}&=&3.830016\left(1+1.120594\Theta+2.078258\Theta^{2}-
80.748072\Theta^{3}\right.\nonumber\\
&+&\left.
  1548.250996\Theta^{4}+0.800424\tau\Theta+
1.183073\tau\Theta^{2}\right).\nonumber
\end{eqnarray}
It can be readily shown that, as expected, $x_{0}$ monotonically 
increases with $\tau$ and $\Theta$.

\subsection{The Kinematic Component}

In an idealized cold IC gas all electrons have the same velocity 
in the CMB frame -- the cluster peculiar velocity, $\beta_{c}$. 
Substituting $\beta_{c}$ for $\beta$ in equation (9), and applying 
the Lorentz transformation
\begin{eqnarray}
\mu_{0}=\frac{\mu_{c}-\beta_{c}}{1-\beta_{c}\mu_{c}}
\label{mu0}
\end{eqnarray}
where $\mu_{c}$ is the angle cosine with respect to the los of an 
observer (as measured in the CMB frame), we obtain
\begin{eqnarray}
\Delta n&=&\frac{3}{32}\tau\left(1-\beta_{c}\mu_{c}\right)\nonumber\\
&\times&\int\left
[\left(3-\mu'^{2}_{0}\right)+\left(3\mu'^{2}_{0}-1\right)\left
(\frac{\mu_{c}-\beta_{c}}{1-\beta_{c}\mu_{c}}\right)^{2}\right]
\delta n d\mu'_{0},
\label{32}
\end{eqnarray}
where $\delta n$ is given in eq. (10), and 
\begin{eqnarray}
z=\frac{1+\beta\mu'_{0}}{1+\beta\mu_{0}}=\gamma_{c}^{2}\left(1+
\beta_{c}\mu'_{0}\right)\left(1-\beta_{c}\mu_{c}\right).
\label{33}
\end{eqnarray}
We evaluate the latter expression for $\Delta n$ as part of the 
calculation of the full effect in the next section.

\subsection{The Full Effect}

To first approximation, when $\Theta$ and $\beta_{c}$ are 
sufficiently small, the impact on the CMB by Compton scattering in 
a cluster with finite peculiar velocity is simply a sum of the 
thermal and kinematic components. For higher values of $\Theta$ 
and $\beta_{c}$, higher order terms have to be included. As a 
result, cross terms linking the two components arise. The 
inclusion of the peculiar velocity of the cluster in a consistent
calculation (not as a separate effect) was originally considered 
by Sazonov \& Sunyaev (1998), and later also by Nozawa et al. 
(1998b). Their calculations yield 
consistent results for the cross terms that depend on both 
$\Theta$ and $\beta_{c}$. We repeat the calculation of the 
full effect by generalizing our treatment described in the 
previous sections. We do so in part because our calculation 
yields slightly different results for the cross terms, although 
our results for the pure thermal and kinematic components are 
exactly identical to those obtained in the latter papers. 

We begin with the transformation of the electron thermal energy 
and momentum, $E_{t}$ and $\vec{p}_{t}$, from the cluster to the 
CMB frame, $E$ and $\vec{p}$, respectively 
\begin{eqnarray}
E &=&\gamma_{c}\left(E_{t}+c\vec{\beta}_{c}\cdot
\vec{p}_{t}\right)\nonumber\\
\vec{p} &=&\gamma_{c}\left(\vec{p}_{t}+\vec{\beta}_{c}E_{t}/c\right).
\label{34}
\end{eqnarray}
Applying these transformations in the expression for the Boltzmann 
factor we obtain
\begin{eqnarray}
\exp\left(-\frac{E_{t}-m_{e}c^{2}}{kT_{e}}\right)=
\exp\left(-\frac{\gamma_{t}-1}{\Theta}\right)\nonumber\\
=\exp\left(-\frac{\gamma_{c}\gamma(1-\vec{\beta}\cdot\vec
{\beta}_{c})-1}{\Theta}\right),
\label{boostedboltzmann}
\end{eqnarray}
where $\beta$ and $\gamma$ are the velocity and Lorentz factor in 
the CMB frame. 

Since $d^{3}p/E$ is invariant under Lorentz transformations, 
\begin{eqnarray}
d^{3}p_{t}&=&\frac{E_{t}}{E}d^{3}p=\gamma_{c}\left(1-\vec{\beta}_{c}\cdot\vec{\beta}\right)p^{2}dpd\mu
d\varphi\nonumber\\
&=&\gamma_{c}\left(1-\vec{\beta}_{c}\cdot\vec{\beta}\right)\gamma^{5}\beta^{2}d\beta
d\mu d\varphi.
\label{37}
\end{eqnarray}
Expanding to second order in $\beta_{c}$ we obtain
\begin{eqnarray}
&
&p_{t}^{2}dp_{t}\exp\left(-\frac{\gamma_{t}-1}{\Theta}\right)=\nonumber\\
&=&\gamma_{c}\left(1-\vec{\beta}_{c}\cdot\vec{\beta}_{t}\right)\exp\left(-\frac{\gamma_{c}\gamma\left(1-\vec{\beta}\cdot\vec{\beta}_{c}\right)-1}{\Theta}\right)\gamma^{5}\beta^{2}d\beta\approx\nonumber\\
&\approx
&\left[1+\vec{\beta}_{c}\cdot\vec{\beta}\left(\frac{\gamma}{\Theta}-1\right)+\left[\frac{\beta_{c}^{2}}{2}\left(1-\frac{\gamma}{\Theta}\right)+\left(\vec{\beta}_{c}\cdot\vec{\beta}\right)^{2}\left(\frac{\gamma^{2}}{2\Theta^{2}}-\frac{\gamma}{\Theta}\right)\right]\right]\nonumber\\
&\times &\exp\left(-\frac{\gamma-1}{\Theta}\right)\gamma^{5}\beta^{2}d\beta.
\label{38}
\end{eqnarray}
Note that
\begin{eqnarray}
\vec{\beta}_{c}\cdot\vec{\beta}&=&\beta_{c}\beta\left(\cos\theta\cos\theta_{c}+\cos\varphi\sin\theta\sin\theta_{c}\right)\nonumber\\
\cos\theta&=&\frac{\cos\theta_{0}+\beta}{1+\beta\cos\theta_{0}}
\label{39}
\end{eqnarray}
where $\theta$ and $\theta_{c}$ are angles in the cosmic frame and
$\theta_{0}$ is in the electron rest frame; we select axes such 
that $\varphi_{c}=0$). Integrating over the azimuthal angle 
$\varphi$ (divided by $2\pi$) and expanding again near 
$\beta_{c}=0$, we obtain 
\begin{eqnarray}
& &p_{t}^{2}dp_{t}\exp\left(-\frac{\gamma_{t}-1}{\Theta}\right)\approx\nonumber\\
&
&\left[1+\beta_{c}\beta\mu_{c}\mu\left(\frac{\gamma}{\Theta}-1\right)+\frac{1}{2}\beta_{c}^{2}\beta^{2}\left(\mu_{c}^{2}-\frac{1}{3}\right)\left(3\mu^{2}-1\right)\left(\frac{1}{2}\frac{\gamma^{2}}{\Theta^{2}}-\frac{\gamma}{\Theta}\right)\right.\nonumber\\
&+&\left.\beta_{c}^{2}\left(\frac{\beta^{2}\gamma^{2}}{6\Theta^{2}}-\frac{\beta^{2}\gamma}{3\Theta}+\frac{1}{2}-\frac{\gamma}{2\Theta}\right)\right]\exp\left(-\frac{\gamma-1}{\Theta}\right)\gamma^{5}\beta^{2}d\beta.
\label{40}
\end{eqnarray}
Substituting this into the transfer equation (Equation (\ref{21})) 
and performing the integrations yield the correction to the 
thermal effect, Equation (\ref{deltantau})
\begin{eqnarray}
\Delta n/\tau&=&-\beta_{c}\mu_{c}\left(\frac{1}{2}A_{1}+\Theta
  F_{13}+\Theta^{2}F_{14}+\Theta^{3}F_{15}+
\Theta^{4}F_{16}\right)\nonumber\\
&+&\beta_{c}^{2}P_{2}\left(\mu_{c}\right)\left(F_{17}+\Theta
  F_{18}+\Theta^{2}F_{19}+\Theta^{3}F_{20}+
\Theta^{4}F_{21}\right)\nonumber\\
&+&\beta_{c}^{2}\left(\frac{1}{3}F_{1}+\Theta F_{22}+
\Theta^{2} F_{23}+\Theta^{3} F_{24}+\Theta^{4} F_{25}\right),
\label{41}
\end{eqnarray}
where $F_{i}$ are found in Appendix A. 

The purely kinematic terms in equation (\ref{41}) 
are identical to the corresponding terms in similar equations 
obtained by Sazonov \& Sunyaev (1998) and Nozawa et al. (1998b). 
However, our results for the cross terms differ somewhat from 
those obtained in the latter two papers.
The difference is due to integration of the transfer 
equation over $\gamma^{5}\beta^{2}d\beta d\mu d\varphi$, rather than over 
equation (\ref{37}) -- as we have done here -- which includes the 
extra factor $\gamma_{c}(1-\vec{\beta}_{c}\cdot\vec{\beta})$. 
From a practical point of view this difference is quite small, 
typically less than 1\%. We note that both our result, equation 
(\ref{41}), and the results derived in all the latter three 
papers satisfy photon number conservation.

\section{Discussion}

In this paper we have first shown that the three main, seemingly 
different approaches to the calculation of relativistic 
corrections to the expressions of Sunyaev \& Zeldovich (1972), 
are indeed equivalent in the limiting case for which they 
are derived. We then extended the accuracy of the analytic 
approximation to the thermal and kinematic terms such that the 
purely thermal component of the effect is accurate to terms 
of order $\Theta^8$, and the cross terms to order 
$\beta_{c}^{2}\Theta^4$. We have also obtained a relatively 
compact fit to the exact numerical calculation of $\Delta n$ 
that is accurate to within 2.5\% for gas temperatures as high 
as 50 keV. Either this functional fit, or the more accurate 
-- but perhaps somewhat less convenient -- fully numerical 
tabulation of Itoh \& Nozawa (2003), can be used in the 
analysis of SZ measurements -- including cosmological parameter 
determination -- whose current observational errors are well 
above the accuracies provided by these fits.

While the accuracy of these expressions is limited very close 
to the crossover frequency, this is usually of little practical 
consequence because of the typically broad detector spectral 
response: Even when the measurements are made very close to the 
crossover frequency for the purpose of measuring cluster 
peculiar velocities, what is required is only a sufficiently 
accurate value of the $\Delta I$ convolved over the spectral 
response. The analytic expression given here is accurate 
to within a few \%. Convolving both the exact (eq. 21) 
and the approximate expressions (eqs. 23 \& 30) over a 
Gaussian detector response centered at 214 GHz, and with a 
FWHM of 30 GHz (characteristics of the second MITO channel), 
we find that the two results agree to within 0.2\%. Thus, the 
impact of this small difference (close to the crossover 
frequency) on the determination of cluster peculiar velocities 
is expected to be negligible. 

The accuracy of the theoretical description of the SZ effect 
surpasses the precision with which the effect can be measured at 
present. In addition to observational errors, systematic modeling 
uncertainties are relatively large. One of these uncertainties 
stems from the unknown gas temperature profile. From a theoretical 
point of view this can be readily assessed for a given assumed 
profile. For example, consider the case of an idealized (and 
perhaps somewhat unrealistic) polytropic equation of state $P 
\propto \rho^{\gamma}$, where $\gamma$ is a free parameter which 
is determined from X-ray measurements. In order to generalize 
our result for the analytic approximation for $\Delta n$ to the 
case of non-isothermal gas, we need to replace the spatially 
dependent parameters by their appropriately weighted expressions. 
Consider, for example, the commonly used $\beta$ model for the 
gas density, $n_{e}(r) \propto [1+(r/r_{c})^2]^{-3\beta/2}$, 
where $r_{c}$ is the gas core radius. (Note that here $\gamma$ 
and $\beta$ denote different quantities than those in the rest 
of the paper.) Since $T_{e}\propto\rho^{1-\gamma}$, the products 
of $\tau \Theta^i$ that appear in the analytic approximation for 
$\Delta n$ have to be replaced by weighted averages along the los
\begin{eqnarray}
\bar{\tau\Theta^{i}}=\tau_{0}\Theta_{0}^{i}\int_{-z}^{z}(1+
x^2)^{-\frac{3}{2}\beta\left[1+\left(\gamma-1\right)i\right]}dx
\end{eqnarray}
where $z$ is the radial extent of the gas distribution in units 
of the core radius, $n_0$ \& $T_0$ are the central values of the 
electron density and temperature, and 
$\tau_{0}=2\sigma_{T} n_{0}r_{c}$.

\section*{Appendix A}

The following is a list of the functions $F_{i}$, $G_{i}$ and $H_{i}$
in terms of the functions $A_{j}$ [defined in Equation (\ref{An})]
\begin{eqnarray}
F_{1}&=&2A_{1}+\frac{1}{4}A_{2}\nonumber\\
F_{2}&=&5A_{1}+\frac{47}{8}A_{2}+\frac{21}{20}A_{3}+\frac{7}{160}A_{4}\nonumber\\
F_{3}&=&\frac{15}{4}A_{1}+\frac{1023}{32}A_{2}+\frac{217}{10}A_{3}+\frac{329}{80}A_{4}+\frac{11}{40}A_{5}+\frac{11}{1920}A_{6}\nonumber\\
F_{4}&=&-\frac{15}{4}A_{1}+\frac{2505}{32}A_{2}+\frac{3549}{20}A_{3}+\frac{14253}{160}A_{4}+\frac{9297}{560}A_{5}\nonumber\\
&+&\frac{12059}{8960}A_{6}+\frac{1}{21}A_{7}+\frac{1}{1680}A_{8}\nonumber\\
F_{5}&=&\frac{135}{64}A_{1}+\frac{30375}{512}A_{2}+\frac{62391}{80}A_{3}+\frac{614727}{640}A_{4}+\frac{124389}{320}A_{5}\nonumber\\
&+&\frac{355703}{5120}A_{6}+\frac{2071}{336}A_{7}+\frac{1879}{6720}A_{8}+\frac{11}{1792}A_{9}+\frac{11}{215040}A_{10}\nonumber\\
F_{6}&=&\frac{45}{16}A_{1}-\frac{7515}{128}A_{2}+\frac{28917}{16}A_{3}+\frac{795429}{128}A_{4}+\frac{2319993}{448}A_{5}\nonumber\\
&+&\frac{12667283}{7168}A_{6}+\frac{201631}{672}A_{7}+\frac{10655}{384}A_{8}+\frac{46679}{32256}A_{9}+\frac{10853}{258048}A_{10}\nonumber\\
&+&\frac{29}{46080}A_{11}+\frac{29}{7741440}A_{12}\nonumber\\
F_{7}&=&-\frac{7425}{512}A_{1}+\frac{128655}{4096}A_{2}+\frac{360675}{256}A_{3}+\frac{50853555}{2048}A_{4}+\frac{45719721}{1024}A_{5}\nonumber\\
&+&\frac{458203107}{16384}A_{6}+\frac{22251961}{2688}A_{7}+\frac{71548297}{53760}A_{8}+\frac{26865067}{215040}A_{9}\nonumber\\
&+&\frac{7313155}{1032192}A_{10}+\frac{1123}{4608}A_{11}+\frac{6361}{1290240}A_{12}+\frac{37}{691200}A_{13}+\frac{37}{154828800}A_{14}\nonumber\\
F_{8}&=&\frac{675}{16}A_{1}+\frac{6345}{128}A_{2}-\frac{86751}{64}A_{3}+\frac{28579473}{512}A_{4}+\frac{463090581}{1792}A_{5}\nonumber\\
&+&\frac{8680356807}{28672}A_{6}+\frac{407333911}{2688}A_{7}+\frac{304758409}{7680}A_{8}+\frac{14281971623}{2365440}A_{9}\nonumber\\
&+&\frac{32154229291}{56770560}A_{10}+\frac{17138321}{506880}A_{11}+\frac{36841447}{28385280}A_{12}+\frac{481771}{15206400}A_{13}\nonumber\\
&+&\frac{229693}{486604800}A_{14}+\frac{23}{5913600}A_{15}+\frac{23}{1703116800}A_{16}\nonumber\\
F_{9}&=&4A_{1}+\frac{17}{4}A_{2}+\frac{3}{4}A_{3}+\frac{1}{32}A_{4}\nonumber\\
F_{10}&=&20A_{1}+\frac{295}{4}A_{2}+\frac{873}{20}A_{3}+\frac{1271}{160}A_{4}+\frac{21}{40}A_{5}+\frac{7}{640}A_{6}\nonumber\\
F_{11}&=&40A_{1}+599A_{2}+\frac{44643}{50}A_{3}+\frac{19557}{50}A_{4}+\frac{27549}{400}A_{5}+\frac{34873}{6400}A_{6}\nonumber\\
&+&\frac{367}{1920}A_{7}+\frac{367}{153600}A_{8}\nonumber\\
F_{12}&=&\frac{45}{2}A_{1}+\frac{96651}{32}A_{2}+\frac{8659449}{800}A_{3}+\frac{62384943}{6400}A_{4}+\frac{38586081}{11200}A_{5}\nonumber\\
&+&\frac{103117227}{179200}A_{6}+\frac{82813}{1680}A_{7}+\frac{590831}{268800}A_{8}+\frac{859}{17920}A_{9}+\frac{859}{2150400}A_{10}\nonumber\\
F_{13}&=&\frac{9}{2}A_{1}+\frac{47}{20}A_{2}+\frac{7}{40}A_{3}\nonumber\\
F_{14}&=&\frac{35}{4}A_{1}+\frac{1023}{40}A_{2}+\frac{833}{80}A_{3}+\frac{183}{160}A_{4}+\frac{11}{320}A_{5}\nonumber\\
F_{15}&=&\frac{175}{16}A_{1}+\frac{18721}{160}A_{2}+\frac{45969}{320}A_{3}+\frac{53757}{1120}A_{4}+\frac{13291}{2240}A_{5}+\frac{3917}{13440}A_{6}\nonumber\\
&+&\frac{1}{210}A_{7}\nonumber\\
F_{16}&=&-\frac{315}{16}A_{1}+\frac{45231}{160}A_{2}+\frac{324387}{320}A_{3}+\frac{355815}{448}A_{4}+\frac{1020813}{4480}A_{5}\nonumber\\
&+&\frac{781943}{26880}A_{6}+\frac{1983}{1120}A_{7}+\frac{167}{3360}A_{8}+\frac{11}{21504}A_{9}\nonumber\\
F_{17}&=&\frac{1}{3}A_{1}+\frac{11}{120}A_{2}\nonumber\\
F_{18}&=&\frac{4}{3}A_{1}+\frac{169}{60}A_{2}+\frac{3}{4}A_{3}+\frac{19}{480}A_{4}\nonumber\\
F_{19}&=&\frac{10}{3}A_{1}+\frac{2609}{120}A_{2}+\frac{2349}{120}A_{3}+\frac{31277}{6720}A_{4}+\frac{1227}{3360}A_{5}+\frac{23}{2688}A_{6}\nonumber\\
F_{20}&=&\frac{5}{2}A_{1}+\frac{14957}{160}A_{2}+\frac{34113}{160}A_{3}+\frac{223653}{1792}A_{4}+\frac{14963}{560}A_{5}+\frac{16249}{6720}A_{6}\nonumber\\
&+&\frac{941}{10080}A_{7}+\frac{101}{80640}A_{8}\nonumber\\
F_{21}&=&-\frac{205}{64}A_{1}+\frac{81869}{512}A_{2}+\frac{6291}{160}A_{3}-\frac{6482541}{8960}A_{4}-\frac{2372827}{4480}A_{5}\nonumber\\
&-&\frac{6820873}{53760}A_{6}-\frac{269683}{20160}A_{7}-\frac{109183}{161280}A_{8}-\frac{515}{32256}A_{9}-\frac{1}{7168}A_{10}\nonumber\\
F_{22}&=&\frac{8}{3}A_{1}+\frac{23}{6}A_{2}+\frac{7}{10}A_{3}+\frac{7}{240}A_{4}\nonumber\\
F_{23}&=&\frac{20}{3}A_{1}+\frac{367}{12}A_{2}+\frac{427}{20}A_{3}+\frac{1967}{480}A_{4}+\frac{11}{40}A_{5}+\frac{11}{1920}A_{6}\nonumber\\
F_{24}&=&5A_{1}+\frac{2131}{16}A_{2}+\frac{19369}{80}A_{3}+\frac{76163}{640}A_{4}+\frac{3099}{140}A_{5}+\frac{12059}{6720}A_{6}\nonumber\\
&+&\frac{4}{63}A_{7}+\frac{1}{1260}A_{8}\nonumber\\
F_{25}&=&-5A_{1}+\frac{4925}{16}A_{2}+\frac{126903}{80}A_{3}+\frac{1085261}{640}A_{4}+\frac{185183}{280}A_{5}\nonumber\\
&+&\frac{522981}{4480}A_{6}+\frac{10387}{1008}A_{7}+\frac{9403}{20160}A_{8}+\frac{55}{5376}A_{9}+\frac{11}{129024}A_{10}\nonumber\\
H_{1}&=&2A_{1}+\frac{1}{4}A_{2}\nonumber\\
H_{2}&=&4.6699698A_{1}+5.4022055A_{2}+0.77091691A_{3}-0.040304645A_{4}\nonumber\\
&-&0.0138114993A_{5}-0.0012957652A_{6}-7.168369\times
10^{-5}A_{7}\nonumber\\
&-&2.3560827\times 10^{-6}A_{8}-4.333988\times 10^{-8}A_{9}-3.4497085\times 10^{-10}A_{10}\nonumber\\
H_{3}&=&19.0998386A_{1}+57.853156A_{2}+39.0362A_{3}+9.6917067A_{4}\nonumber\\
&+&1.206357A_{5}+0.09119917A_{6}+0.004517831A_{7}+1.4162735\times
10^{-4}A_{8}\nonumber\\
&+&2.5544641\times 10^{-6}A_{9}+2.0823368\times 10^{-8}A_{10}\nonumber\\
H_{4}&=&349.87908A_{1}+249.46964A_{2}+105.99813A_{3}+41.012553A_{4}\nonumber\\
&+&8.7994039A_{5}+0.96698562A_{6}+0.058225778A_{7}+0.0019859754A_{8}\nonumber\\
&+&3.59624\times 10^{-5}A_{9}+2.6060166\times 10^{-7}A_{10}\nonumber\\
H_{5}&=&-2818.9102A_{1}-3109.3757A_{2}-1577.9854A_{3}-465.38533A_{4}\nonumber\\
&-&79.009596A_{5}-7.7208374A_{6}-0.44258019A_{7}-0.01482893A_{8}\nonumber\\
&-&2.6904478\times 10^{-4}A_{9}-2.0134634\times 10^{-6}A_{10}\nonumber
\end{eqnarray}
\begin{flushright}
(A.1)
\end{flushright}

\section*{Appendix B}

From Equations (22) and (24) the following relations are obtained 
\begin{eqnarray}
A_{1}\sinh\left(\frac{x}{2}\right)&=&-\frac{1}{2}\tilde{S}\nonumber\\
A_{2}\sinh\left(\frac{x}{2}\right)&=&\tilde{S}\tilde{X}\nonumber\\
A_{3}\sinh\left(\frac{x}{2}\right)&=&-2\tilde{S}\tilde{X}^{2}-\tilde{S}^{3}\nonumber\\
A_{4}\sinh\left(\frac{x}{2}\right)&=&4\tilde{S}\tilde{X}^{3}+8\tilde{S}^{3}\tilde{X}\nonumber\\
A_{5}\sinh\left(\frac{x}{2}\right)&=&-8\tilde{S}\tilde{X}^{4}-44\tilde{S}^{3}\tilde{X}^{2}-8\tilde{S}^{5}\nonumber\\
A_{6}\sinh\left(\frac{x}{2}\right)&=&16\tilde{S}\tilde{X}^{5}+208\tilde{S}^{3}\tilde{X}^{3}+136\tilde{S}^{5}\tilde{X}\nonumber\\
A_{7}\sinh\left(\frac{x}{2}\right)&=&-32\tilde{S}\tilde{X}^{6}-912\tilde{S}^{3}\tilde{X}^{4}-1440\tilde{S}^{5}\tilde{X}^{2}-136\tilde{S}^{7}\nonumber\\
A_{8}\sinh\left(\frac{x}{2}\right)&=&64\tilde{S}\tilde{X}^{7}+3840\tilde{S}^{3}\tilde{X}^{5}+12288\tilde{S}^{5}\tilde{X}^{3}+3968\tilde{S}^{7}\tilde{X}\nonumber\\
A_{9}\sinh\left(\frac{x}{2}\right)&=&-128\tilde{S}\tilde{X}^{8}-15808\tilde{S}^{3}\tilde{X}^{6}-92928\tilde{S}^{5}\tilde{X}^{4}-68608\tilde{S}^{7}\tilde{X}^{2}\nonumber\\
&-&3968\tilde{S}^{9}\nonumber\\
A_{10}\sinh\left(\frac{x}{2}\right)&=&256\tilde{S}\tilde{X}^{9}+64256\tilde{S}^{3}\tilde{X}^{7}+652416\tilde{S}^{5}\tilde{X}^{5}+920576\tilde{S}^{7}\tilde{X}^{3}\nonumber\\
&+&176896\tilde{S}^{9}\tilde{X}\nonumber.
\end{eqnarray}
\begin{flushright}
(B.1)
\end{flushright}

\begin{figure}[h]
\begin{center}
\includegraphics*[width=10cm]{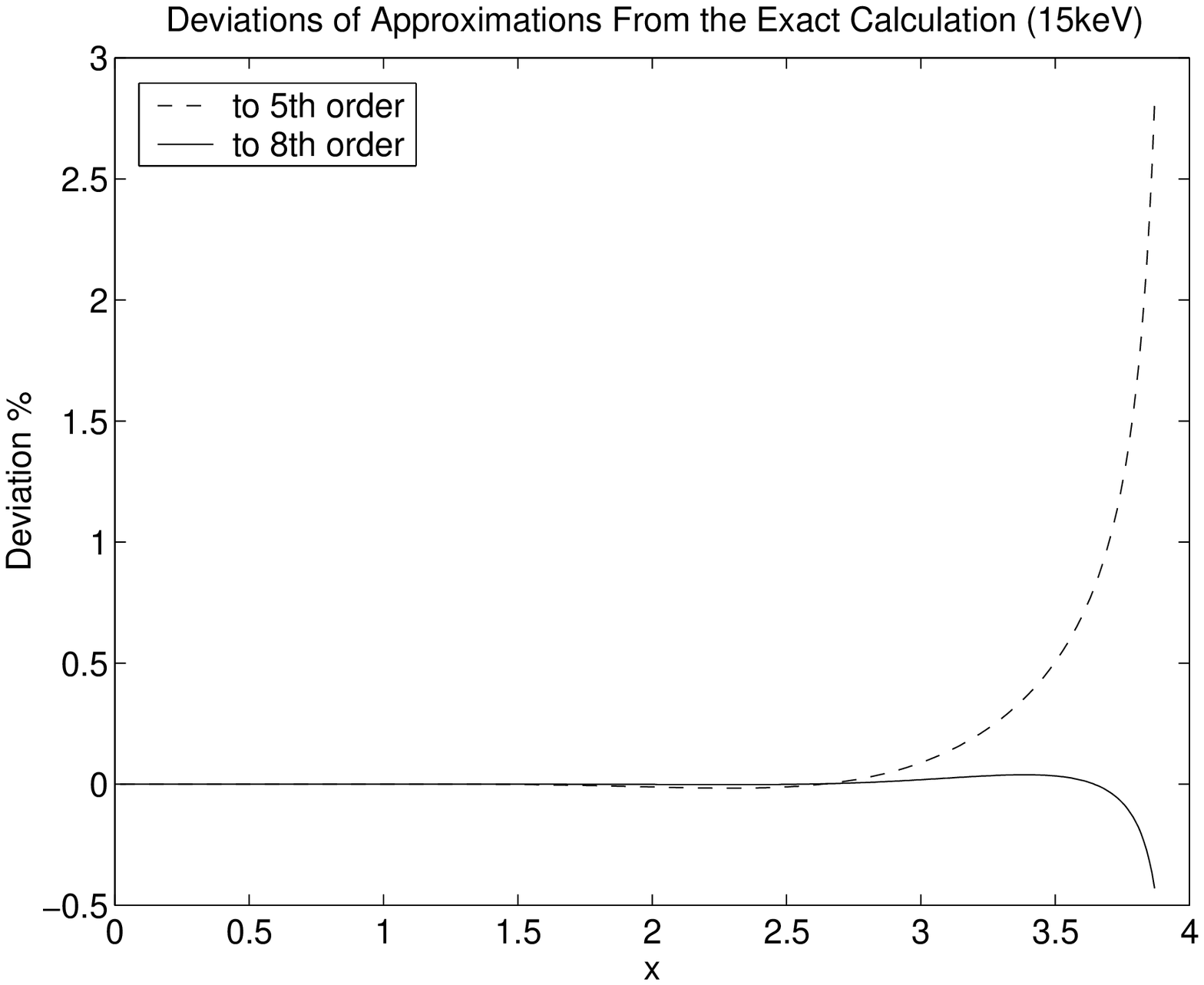}
\includegraphics*[width=10cm]{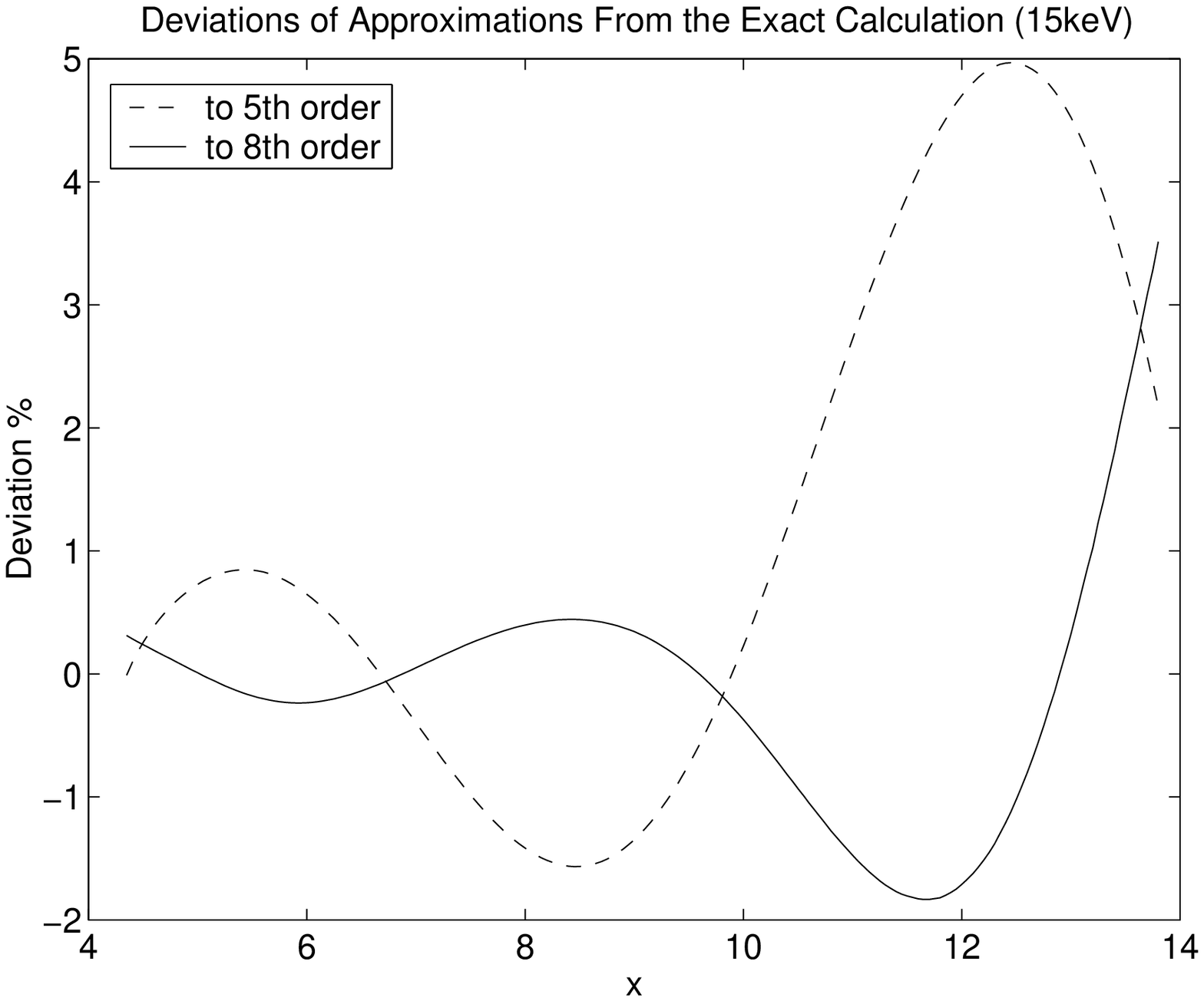}
\end{center}
\caption{The deviations of the fifth (Itoh et al. 2000a) and 
eighth order approximations (equation 23) from the exact result.}
\end{figure}

\end{document}